 \documentclass[
a4paper,article, preprint, amsmath, amssymb, floatfix,
nofootinbib,wrapfloat byrevtex]{revtex4}

\usepackage{lineno,hyperref}
\modulolinenumbers[5]

 \usepackage{amsmath,mathrsfs,tabularx,graphicx,epstopdf,placeins,float}
 \usepackage{booktabs}
 \usepackage{multirow}
 \usepackage[titletoc]{appendix}
 \usepackage{titlesec}
 \usepackage{caption}
 \usepackage[labelformat=empty]{caption}
 \usepackage{color}
\usepackage{changepage}
\usepackage{graphicx}
\usepackage{subfig}
\newcolumntype{C}{>{\centering\arraybackslash}X}
\usepackage[margin=2.5cm]{geometry}

 \newcommand{\beq}[1]{\begin{equation}\label{#1}}
 \newcommand{\eeq}{\end{equation}}
 \newcommand{\bea}[1]{\begin{eqnarray}\label{#1}}
 \newcommand{\eea}{\end{eqnarray}}
 \newcommand\figcaption{\def\@captype{figure}\caption}
 \newcommand\tabcaption{\def\@captype{table}\caption}
 
\begin{document}
 \title{Constrained dynamics of maximally entangled bipartite system}
 \author{Asma Bashir$^{1}$}
\email{kbasmabashir807@gmail.com}
\author{Muhammad Abdul Wasay$^{1}$}
\email{wasay31@gmail.com}
\affiliation{$^1$Department of Physics, University of Agriculture, Faisalabad 38040, Pakistan.}
 \begin{abstract}
 The classical and quantum dynamics of two particles constrained on $S^1$ is discussed via Dirac's approach. We show that when state is maximally entangled between two subsystems, the product of dispersion in the measurement reduces. We also quantify the upper bound
 on the external field $\vec{B}$ such that $\vec{B}\geq\vec{B}_{\textmd{upper}}$ implies no reduction in the product of dispersion pertaining to one subsystem. Further, we report on the cut-off value of the external field $\vec{B}_{\textmd{cutoff}}$, above which the bipartite entanglement is lost and there exists a direct relationship between uncertainty of the composite system and the external field. We note that, in this framework it is possible to tune the external field for entanglement/unentanglement of a bipartite system. Finally, we show that the additional terms arising in the quantum Hamiltonian, due to the requirement of Hermiticity of operators, produce a shift in the energy of the system.
  \end{abstract}

 \maketitle
 \smallskip

\section{INTRODUCTION}
\par From a classical perspective, all dynamical variables commute but in quantum mechanics the uncertainty relations are fundamental. The seminal work of EPR \cite{EPR} reports that if two systems are correlated and the physical quantities pertaining to one system are determined then uncertainty in their prediction for the second system is zero. The compatibility between any two observables $\hat{A}$ and $\hat{B}$ is usually understood in terms of the Robertson relation \cite{Robertson}
\bea{}
(\Delta \hat{A})^2(\Delta \hat{B})^2\geq \frac{|\langle \hat{A},\hat{B} \rangle|^2}{4}
\label{Robertson}
\eea
where the dispersion in the measurement reduces for small lower bounds.
On the other hand, observables over a bipartite quantum state were analysed for uncertainty in their measurement by evaluating the upper bound in terms of entropies\cite{Berta PRA}, and it was observed that uncertainty in the predictions is reduced if both parties are maximally entangled and they strongly agree on the measurement. Thus a measure of uncertainty in the prediction can be obtained by a measure of the entanglement between the systems. Quantum entanglement and uncertainty relations are therefore strongly related
and have wide applications in quantum information and computing \cite{A.stean,Bennett} and so on.  An effort to study the effect of quantum correlation on the uncertainty was made in ref.\cite{Rigolin} which presented a Generalized Uncertainty Principle (GUP) for bipartite and tripartite entangled states. It was observed that the lower bound in GUP deviates from the Heisenberg Uncertainty Principle (HUP) and leads to a reduced uncertainty for quantum correlated particles. It is interesting to explore these issues in the framework of constrained quantum theory.
\par  The Hamiltonian formulation of systems in the presence of first and second class constraints has been reported in literature extensively since the inception of the idea \cite{Dirac}. Ref.\cite{PLA} presents a similar study for the dynamics of a single particle constrained on a circle. The Hamiltonian formulation is obtained from Lagrangian mechanics as $p_i=\partial L/\partial\dot{x}_i$. Expressing the velocities this way makes the Hamiltonian to be independent of $\dot{x}_i$ and the Hessian of Lagrangian is given by
\bea{}
\left|\left|\frac{\partial^2L}{\partial \dot{x}_i \partial\dot{x}_j}\right|\right|\neq 0
\label{Hessian}
\eea
The equality holds for the case of degenerate Lagrangian where the constrained dynamics of the system can not be obtained in usual way by evaluating the Poisson brackets but by Hamilton's equations that contain Dirac brackets \cite{Dirac}\cite{PLA}.

\par Refs. \cite{Vaisman}-\cite{pavlov} deal with the problem of assigning geometric meanings to the Hamiltonian dynamics or designing of a proper quantum framework for the dynamics of particle on curved spaces. Dirac's approach for handling these problems is presented in \cite{pavlov}, where the dynamics in the presence of second class constraints is described by defining a symplectic structure. A similar study in this direction was made in \cite{D.M.Xun} for the Dirac quantization of particle constrained on a helicoid and in \cite{Kleinert} for quantizing the dynamics of a free particle on a $D$-dimensional sphere. A  different approach to dualize quantum with geometry in the pilot-wave limit is presented in \cite{Asma,Wasay1,Wasay3}, and with topological properties in \cite{Wasay2}.
Dirac quantization on curved spaces adds some additional terms in the Schr\"{o}dinger equation that can be linked with the curvature of the space or distortion in the energy spectrum. It will be interesting to discuss such energy terms for two maximally entangled subsystems, in connection with the particles constrained on curved spaces \cite{KS cheng}.
\par
The purpose of this paper is to study a bipartite quantum system in which the degrees of freedom of the system are reduced by constraining them on $S^1$. For this, we consider an inseparable spatial wavefunction like the one considered in ref.\cite{Wasay3}.
\bea{}
\phi(x_1,x_2)\neq \phi(x_1)\phi(x_2)
\label{productstate}
\eea
the fundamental criterion that the composite state for entangled particles cannot be written as a product state but as a superposition of product states with concurrence $C$ \cite{Schroeder}-\cite{Dalton} as
\bea{}
|\Phi\rangle=\sum_{ijk...}A_{ijk}|\phi_a^i\rangle\otimes|\phi_b^j\rangle\otimes|\phi_c^k\rangle...
\label{entangledstate}
\eea
Taking such a two particle composite state we will analyse the compatibility between position and momentum operators.
We will establish Hermitian expression for momentum operators and finally we  will comment on the energy terms arising in Schr\"{o}dinger equation in connection with the quantization on curved manifolds \cite{KS cheng}
\par
The paper is organized as: Section-II deals with studying the classical dynamics of two particles constrained on $S^1$. In section-III, we present Dirac quantization of the classical dynamics assuming the two particles as two maximally entangled subsystems and comment on the uncertainty between observables. In section-IV, we establish Hermitian expression for momentum operators and find the energy equation for two constrained entangled particles. Section-V summarizes the results.

\section{Constrained classical dynamics of two particles on $S^1$}

We consider a two particle Lagrangian in scalar and vector potential given by
\bea{}
L=\sum_k \left[\frac{1}{2}m \dot{r}_k^2-eV(\vec{r},t)+e\dot{r}_k\cdot \vec{A}_{r_k}(\vec{r},t)\right]
\label{L1}
\eea
where $V$ is the scalar potential, $\vec{A}$ is the vector potential, $e$ and $m$ are the charge and mass of the particle respectively and $\dot{r}_k$ represents the particle velocity, $k=1,2$ is the particle index
and $\dot{r}_k^2=\dot{r}_1^2+\dot{r}_2^2=\dot{x}_1^2+\dot{y}_1^2+\dot{x}_2^2+\dot{y}_2^2$. We assume these particles to be under the effect of a uniform external magnetic field $\vec{B}$ directed along the normal to the plane containing them and as a result, these particles retain their motion on $S^1$ i.e., $g(x,y)=x_k^2+y_k^2-r_k^2$. Applying the constraint $g(x,y)=0$ on the system, the resulting Lagrangian is
\bea{}
L= \sum_k\left[\frac{1}{2} \dot{r}_k^2-eV(\vec{r},t)+e\dot{r}_k\cdot \vec{A}_{r_k}(\vec{r},t)-\lambda (x_k^2+y_k^2-r_k^2)\right]
\label{L2}
\eea
where we have set $m=1$.

With momentum given by $P_{r_k}=\dot{r}_k+e\vec{A}_{r_k}$, the Hamiltonian of the system becomes
\bea{}
H=P_{\lambda}\dot{\lambda}+\sum_k\left[\frac{1}{2} (P_{r_k}-e\vec{A}_{r_k})^2+eV(\vec{r},t)+\lambda (x_k^2+y_k^2-r_k^2)\right]
\label{H1}
\eea

The Hamiltonian in Eq.\eqref{H1} is not independent of velocities and therefore it is not possible to implement the standard approach to obtain equations of motion. Therefore, we need to use Dirac's method for singular Lagrangians \cite{Dirac}. According to this one must generalize the Hamiltonian to incorporate the constraints. This generalized Hamiltonian is obtained by multiplying the primary constraint with an arbitrary function of time, say $u_1$, and adding the resulting term in Hamiltonian Eq.\eqref{H1}
\bea{}
H_T=\sum_k\left[\frac{1}{2} (P_{r_k}-e\vec{A}_{r_k})^2+eV(\vec{r},t)+\lambda (x_k^2+y_k^2-r_k^2)\right]+u_1P_\lambda
\label{H2}
\eea

where $\dot{\lambda}$ is absorbed in $u_1$. One can see that $P_\lambda=\frac{\partial L}{\partial \dot{\lambda}}=0$, which is a primary constraint as it is derived directly from the Lagrangian. We will denote it by $\sigma_1 (=P_\lambda)$.


If a constraint is initially zero, it must be zero for all times. This consistency condition $\dot{\sigma}=[\sigma,H]\approx0$ yields a set of equations\cite{Dirac}. If the resulting equation contains only $x$ or $P$, it is treated as another constraint. The consistency condition on $\sigma_1$ yields an equation of the following form

\bea{}
\sigma_2=x_k^2+y_k^2-2a^2=0
\label{sigma2}
\eea

This equation will serve as the secondary constraint, where $r_k^2=r_1^2+r_2^2=2a^2$. The particles are moving on the same circle so $r_1=r_2=a$, and since $\sigma_2=0$, it must be zero during its evolution in time, which leads to

\bea{}
\sigma_3=r_k\cdot P_{r_k}-er_k\cdot\vec{A}_{r_k}=0
\label{sigma3}
\eea

This equation is of the form $\omega(r,p)=0$ and it will lead to another such consistency condition. So we continue to evaluate these equations until another type of equation (like the one for function $u_1$) is obtained.

\bea{}
\sigma_4=(P_{r_k}-e\vec{A}_{r_k})^2-er_k\nabla_{r_k} V-2\lambda r_k^2=0
\label{sigma4}
\eea

In obtaining the expression for $\sigma_4$, we have used the Coulomb gauge $\vec{\nabla}\cdot \vec{A}=0$.
Now evaluating $\dot{\sigma}_4(r_k, P_{r_k},\lambda, P_\lambda)=[\sigma_4,H_T]=0$ leads to the following expression for the function $u_1$

\bea{}
u_1=-\frac{\left[(P_{r_k}-e\vec{A}_{r_k})\{3e\nabla_{r_k}V+4\lambda r_k+er_k\nabla_{r_k}^2V\}\right]}{2r_k^2}
\label{u1}
\eea

All the constraints obtained are of second class and can be set equal to zero to switch from Poisson brackets to Dirac brackets. The total Hamiltonian \eqref{H2} with the constraint $\sigma_1=\sigma_2=0$ is thus

\bea{}
H_T=\frac{1}{2}(P_{r_k}-e\vec{A}_{r_k})^2+eV
\label{H3}
\eea

For any two functions of canonical variables, say $X$ and $Y$, the Dirac brackets for this system are found following ref.\cite{Dirac}, as

\bea{}
[X,Y]_D=[X,Y]_P-\sum^4_{m,n=1}[X,\sigma_m]_P\Delta_{mn}[\sigma_n,Y]_P
\label{diracbracket}
\eea

where $\sigma_m$ are the primary and secondary constraints described above, and $\Delta_{mn}$ is a matrix such that $\Delta_{mn}=(\Phi^{-1})_{mn}$ which is obtained via the Poisson brackets $\Phi_{mn}=[\sigma_m,\sigma_n]_P$.
The matrix $\Phi_{mn}$ is invertible and spans the subspace of second class constraints and has at least one non vanishing bracket for any two constraints $\sigma_m$ and $\sigma_n$ \cite{PLA}. The inverse matrix $\Delta_{mn}$ is

\begin{gather}
\hspace{-1cm}\Delta_{mn}  =
   \begin{pmatrix}
    0 & \frac{-2(P_{r_k}-eA_{r_k})^2-4\lambda r_k^2-e r_k\nabla_{r_k}V-er_k^2\nabla_{r_k}^2V}{4r_k^4} & \frac{r_k.(P_{r_k}-eA_{r_k})}{r_k^4}& \frac{-1}{2r_k^2} \\
                     \frac{2(P_{r_k}-eA_{r_k})^2+4\lambda r_k^2+e r_k\nabla_{r_k}V+er_k^2\nabla_{r_k}^2V}{4r_k^4} & 0 & \frac{-1}{2r_k^2} & 0\\
                     -\frac{r_k.(P_{r_k}-eA_{r_k})}{r_k^4} & \frac{1}{2r_k^2}  & 0 & 0\\
                     \frac{1}{2r_k^2} & 0 & 0 & 0\\
   \end{pmatrix}\nonumber
\end{gather}


Using Dirac's approach, we arrive at following brackets describing the classical dynamics of two constrained particles.

\bea{}
[x_k,P_{x_k}]_D=\left(1-\frac{x^2_k}{2a^2}\right)=\frac{y_k^2}{2a^2}
\label{xp_x bracket}
\eea
\bea{}
[y_k,P_{y_k}]_D=\left(1-\frac{y^2_k}{2a^2}\right)=\frac{x^2_k}{2a^2}
\label{yp_y bracket}
\eea

where $k=1,2$. The results of these classical brackets are different from those of standard canonical brackets in that the information on the constrained dynamics of the particles is apparent. One can see instead of the usual translation along $x$-axis by $P$ in Eq.\eqref{xp_x bracket}, the motion of particle $1$ is shifted by an amount $\frac{x^2_1}{r_k^2}$ towards $y$-axis. Then Eq.\eqref{yp_y bracket} shifts the motion slightly towards $x$-axis. This series of shifts renders a circular trajectory. The same is true for particle $2$. The Dirac brackets for the coordinates of one particle and momenta of the other particle are given by

\bea{}
[x_j,P_{x_k}]_D=\frac{-x_jx_k}{2a^2}\quad,\quad [y_j,P_{y_k}]_D=\frac{-y_jy_k}{2a^2}
\label{x_jp_xk bracket}
\eea
\bea{}
[x_k,P_{y_k}]_D=\frac{-x_ky_k}{2a^2}\quad,\quad[y_k,P_{x_k}]_D=\frac{-x_ky_k}{2a^2}
\label{xp_y bracket}
\eea
\bea{}
[x_j,P_{y_k}]_D=\frac{-x_jy_k}{2a^2}\quad,\quad[y_j,P_{x_k}]_D=\frac{-x_ky_j}{2a^2}
\label{x_jp_yk bracket}
\eea

 where $j=k=1,2$ and $j\neq k$. Also note that all the brackets in Eqs.\eqref{x_jp_xk bracket}-\eqref{x_jp_yk bracket} vanish in classical mechanics since momentum is not a generator of translations along a direction perpendicular to the particle position. But in Dirac's formalism, the brackets $[x,P_y]=[y,P_x]\ne 0$, which is due to the constraint which forces the particle to remain on $S^1$. The Dirac brackets detailing translations in $P$ space are the following,

\bea{}
[x_j,x_k]_D=[y_j,y_k]_D=[x_k,y_k]_D=[x_j,y_k]_D=0
\label{xy bracket}
\eea

 Since $x$ and $y$ generate translations in momentum space, the brackets in Eq.\eqref{xy bracket} are equivalent to the corresponding Poisson brackets. This is because there is no such $\sigma_m$ whose time evolution results in an equation containing only $P$, there is thus no constraint imposed on the particle's momentum. This is why these results are just analogous to those obtained for a classical Hamiltonian with no constraint on the particle dynamics.
The Dirac brackets for momenta of both the particles are

\bea{}
[P_{x_k},P_{y_k}]_D=\frac{1}{2a^2}\left[-L_z^{(k)}+e(r_k\times \vec{A}_{r_k}).\hat{e}_3\right]
\label{p_xp_y bracket}\eea
\bea{}
[P_{x_j},P_{y_k}]_D=\frac{1}{2a^2}[(y_kP_{x_j}-x_jP_{y_k})+e(x_j\vec{A}_{y_k}-\vec{A}_{x_j}y_k)]
\label{p_xjp_yk bracket}
\eea
\bea{}
[P_{x_1},P_{x_2}]_D=\frac{1}{2a^2}[(x_2P_{x_1}-x_1P_{x_2})+e(x\times \vec{A}_{x}).\hat{e}_3]
\label{p_x1p_x2 bracket}
\eea
\bea{}
[P_{y_1},P_{y_2}]_D=\frac{1}{2a^2}[(y_2P_{y_1}-y_1P_{y_2})+e(y\times \vec{A}_{y}).\hat{e}_3]
\label{p_y1p_y2 bracket}
\eea

where $j=k=1,2$ and $j\neq k$. Eqs. \eqref{p_xp_y bracket}-\eqref{p_y1p_y2 bracket} clearly reflect the rotational dynamics of particle due to explicit presence of $L_z$ or by writing momentum in terms of angular momentum as discussed in the next section. One can see when dealing with degenerate Lagrangian, the Poisson brackets are generalized to Dirac brackets \cite{Dirac}.

\section{DIRAC QUANTIZATION FOR TWO ENTANGLED SUBSYSTEMS}

To quantize the classical dynamics  discussed above, the system with $k=2$ (two particles) is partitioned into two subsystems $1$ and $2$ containing $i$ and $j$ particles respectively, such that $k=i\!+\!j$ with $i\!\!=\!\!j\!\!=\!\!1$, i.e., each particle is treated as a subsystem. The basis for subsystem $1$ are chosen to be $\{\psi_1,\psi_2\}$ corresponding to the Hilbert space $\mathcal{H}_1$ and  $\{\phi_1,\phi_2\}$ for the subsystem $2$ corresponding to the Hilbert space $\mathcal{H}_2$. The composite space for the bipartite quantum state is a tensor product

\bea{}
\mathcal{H}=\mathcal{H}_1\otimes \mathcal{H}_2
\label{hilbertspace1}
\eea

 A pure state with respect to this partitioning can be defined as

\bea{}
\psi(x_1,x_2)=\varepsilon\psi_1(x_1)\phi_1(x_2)+\gamma\psi_1(x_1)\phi_2(x_2)+\eta\psi_2(x_2)\phi_1(x_1)+\delta\psi_2(x_1)\phi_2(x_2)
\label{state}
\eea

One can consider the concurrence as a measure of the entanglement of a pure bipartite quantum state\cite{Lohmayer}, which should be equal to unity for a maximally entangled state and is defined as $C=2\mid\!\varepsilon\delta-\gamma\eta\!\mid$.
%

We choose $\gamma=\eta=0$ and the pure bipartite state is maximally entangled with equal probability of collapse upon measurement, to one of the product state $\psi_1(x_1)\phi_1(x_2)$ or $\psi_2(x_1)\phi_2(x_2)$, i.e., $|\varepsilon|^2+|\delta|^2=1$ \cite{Schroeder}\cite{Dalton}.
The quantum state of two entangled particles on a circle of radius $a$ is chosen to be

\bea{}
\Psi(\chi_1,\chi_2)= e^{i\chi_1\chi_2}(\varphi e^{i(\chi_1+\chi_2)})
\label{polarstate}
\eea

where $\varphi=\sqrt{2}a^2$ and $\chi_k$ gives the angular position of the particles. The classical to quantum transition of the system of two entangled particles is achieved by using the Dirac quantization rule: $\{~,~\}\rightarrow\frac{1}{i\hbar} [~,~]$,
where $\hbar=1$. The following commutation relations for the position and momentum operators are obtained as a result of this canonical quantization.

\bea{}
[\hat{x}_k,\hat{P}_{x_k}]=i\left(1-\frac{\hat{x}^2_k}{2a^2}\right)=\frac{i\hat{y}_k^2}{2a^2}
\label{xp com}
\eea
\bea{}
[\hat{y}_k,\hat{P}_{y_k}]=i\left(1-\frac{\hat{y}^2_k}{2a^2}\right)=\frac{i\hat{x}_k^2}{2a^2}
\label{yp com}
\eea
\bea{}
[\hat{x}_j,\hat{P}_{x_k}]=\frac{-i\hat{x}_j\hat{x}_k}{2a^2}\quad,\quad[\hat{x}_j,\hat{P}_{y_k}]=\frac{-i\hat{x}_j\hat{y}_k}{2a^2}
\label{x_jp_xk com}
\eea
\bea{}
[\hat{x}_k,\hat{P}_{y_k}]=\frac{-i\hat{x}_k\hat{y}_k}{2a^2}\quad,\quad[\hat{y}_k,\hat{P}_{x_k}]=\frac{-i\hat{x}_k\hat{y}_k}{2a^2}
\label{xp_y com}
\eea
\bea{}
[\hat{y}_j,\hat{P}_{y_k}]=\frac{-i\hat{y}_j\hat{y}_k}{2a^2}\quad,\quad[\hat{y}_j,\hat{P}_{x_k}]=\frac{-i\hat{x}_k\hat{y}_j}{2a^2}
\label{y_jp_yk com}
\eea
\bea{}
[x_j,x_k]=[y_j,y_k]=[x_k,y_k]=[x_j,y_k]=0
\label{xy com}
\eea
\par where $j=k=1,2$ and $j\neq k$. The quantum commutation relations in Eqs.\eqref{xp com}-\eqref{xy com} are obtained for the quantum state in Eq.\eqref{polarstate}. From ref.\cite{Rigolin}, we note that for a system of entangled particles, the uncertainty relation between the operators $\hat{x}$ and $\hat{P}$ diverges from the standard HUP in a way that the precision of simultaneous measurement of $\hat{x}$ and $\hat{P}$ is increased. Further note that Eqs.\eqref{xp com}-\eqref{y_jp_yk com} illustrate a reduced uncertainty between position and momentum, in comparison to the results of \cite{PLA}, as expected for maximally entangled subsystems described in \cite{Berta PRA}\cite{Rigolin}.

As an example, one can see that the lower bound evaluated for Eq.\eqref{xp com} using Robertson relation for the parameters values $a^2=10$, with $a^2=x_1^2+y_1^2$ and the choice $x_1=1$, $y_1=3$, gives

\bea{}
(\Delta \hat{x}_1)^2(\Delta \hat{P}_{x_1})^2\geq 0.05
\label{xpbound}
\eea

The product of dispersion reduces in comparison to \cite{PLA} where $(\Delta \hat{x})^2(\Delta \hat{P}_{x})^2\geq 0.2025$ (calculated for the same $a$). Similarly, one can see the lower bounds in the set Eqs.\eqref{xp com}-\eqref{y_jp_yk com} reduces in comparison to \cite{PLA}.

This uncertainty is inversely related to $k$ (the number of particles) such that in a system with several particles ($k\rightarrow\infty$), the standard deviation in the measurement of observables is significantly reduced. In this case, the
uncertainty in the prediction of observables in the set Eqs.\eqref{xp com}-\eqref{y_jp_yk com} can approach zero and in this limit these commutators tend to the Poisson brackets of classical mechanics.

The commutation relation between momentum of each particles is obtained via Dirac quantizing the classical bracket in Eq.\eqref{p_xp_y bracket}, which yields

\bea{}
[\hat{P}_{x_k},\hat{P}_{y_k}]=\frac{i}{2a^2}\left[-\hat{L}_z^{(k)}+e(\hat{r}_k\times \vec{A}_{r_k}).\hat{e}_3\right]
\label{p_xp_y com}
\eea
To comment on Eq.\eqref{p_xp_y com}, we check the Robertson uncertainty relation for momentum of two entangled particles as follows,  
\bea{}
(\Delta \hat{P}_{x_k})^2(\Delta \hat{P}_{y_k})^2\geq \frac{|-(1-\alpha)-\chi_j+e\vec{A}(\hat{x}_k-\hat{y}_k)|^2}{(2k)^2a^4}\left(=1.97\times 10^{-3}\right)
\label{p_xp_ybound}
\eea

We arrive at the numerical result on the r.h.s of Eq.\eqref{p_xp_ybound} with the parameter values: $a^2=10$ so that $2a^2=x_1^2+x_2^2+y_1^2+y_2^2=20$ with $x_1=2$, $x_2=3.1$, $y_1=2.45$, $y_2=0.65$, charge of the particle $e=1$ and the vector potential $A=0.50$.
Similarly, for particle 2, we get

\bea{}
(\Delta \hat{P}_{x_2})^2(\Delta \hat{P}_{y_2})^2\geq 6.5\times 10^{-3}
\label{p_x2p_y2bound}
\eea
The numerical bounds in Eqs.\eqref{p_xp_ybound}-\eqref{p_x2p_y2bound} are computed in view of the single particle case $((\Delta \hat{P}_{x})^2(\Delta \hat{P}_{y})^2\geq  0.0225)$ evaluated for the same $a$.
Clearly, one can see that the lower bound between the momenta of each particle is reduced if: (1). The position of the second particle is known, and: (2). The external field is properly tuned in order to keep the state maximally entangled between particle $1$ and particle $2$. We arrive at these results using a uniform magnetic field of strength $\vec{B}=2A/a=0.32$ Tesla. The correlation is not affected by this field for strengths up to $\vec{B}_{\textmd{upper}}= 1.03$ Tesla and thus the state $\Psi(\chi_1,\chi_2)$ remains inseperable. However, for $\vec{B}\geq\vec{B}_{\textmd{upper}}$, one can see there's no reduction in the uncertainty for particle $2$ and at $ \vec{B}\sim11.26$ Tesla, there is no reduction in uncertainty for both particles. We may denote this field strength as the cut-off, and represent this as $\vec{B}_{\textmd{cutoff}}=11.26$ Tesla.  This is because the interaction between the particles and the field destroys the correlation between both subsystems above this cut-off, which results in the loss of entanglement and the usual uncertainty relation for momentum is restored \cite{PLA}. Hence, for $\vec{B}>\vec{B}_{\textmd{cutoff}}$, there exists a direct relationship between uncertainty of the composite system and the external field.

We can obtain $\vec{B}_{\textmd{upper}}$ of a constrained bipartite entangled system for any $k\geq 2$ with parameters $x_2,y_2$, via equations of the following form
\bea{}
(x_2+n)^2+(y_2+n)^2=r^2
\label{parameters1}
\eea
where $x_2=3.1$, $y_2=0.65$, and $n=\pm0,\pm1,\pm2,\pm3,...$.
The vector potential that gives $\vec{B}_{\textmd{upper}}$ can be obtained as
\bea{}
A_{\textmd{upper}}=A_{0}+p(1.22)
\label{A1}
\eea

with $A_0=1.63$, $p=k-2$ where $k$ is the number of particles, one can see that inserting any value of $k$ in Eq.\eqref{A1} yields the value of $A_{upper}$. Also, from Eqs.\eqref{parameters1}-\eqref{A1}, $A_{upper}$ can be obtained for any $k$ and any $x_2+n$, $y_2+n$ with $n=\pm0,\pm1,\pm2,\pm3,...$ and $x_2=3.1$ and $y_2=0.65$.
We have used the coordinates of subsystem $2$, this is because at $\vec{B}_{\textmd{upper}}$, entanglement reduces due to the interaction between particles and the field and no reduction in the dispersion of observables is possible for subsystem $2$ (however, for subsystem $1$ the lower bound of Robertson relation still reduces).
Similarly, for equations of the following form
\bea{}
(x_1+n)^2+(y_1+n)^2=r^2
\label{parameters2}
\eea
where $x_1=2$ and $y_1=2.45$. The vector potential that leads to $\vec{B}_{\textmd{cutoff}}$, can be written as
\bea{}
A_{\textmd{cutoff}}=\tilde{A}_{0}+p(6.7)
\label{A2}
\eea
where $\tilde{A}_0=17.8$ and we have used the coordinates of subsystem $1$ because at the cut-off value, the entanglement dies and no reduction in the dispersion is possible for subsystem $1$. Eq.\eqref{A2} gives the cut-off value of the external magnetic field such that for $A>A_{\textmd{cutoff}}$, the two subsystems are unentangled and the lower bound of Robertson relation increases with increasing $\vec{B}$ above the cut-off. We arrive at Eq.\eqref{A1} and Eq.\eqref{A2} by inspection i.e., by varying the value of vector potential $A$ and computing the lower bounds in Eqs.\eqref{p_xp_ybound}-\eqref{p_x2p_y2bound}. Proceeding this way we can obtain the value of $A_{cutoff}$ and $A_{upper}$ at which no reduction in the lower bound is obtained for Eq.\eqref{p_xp_ybound} and Eq.\eqref{p_x2p_y2bound} (in view of the single particle case) respectively. However, the choice of subsystem $1$ and $2$ is arbitrary and one only needs to consider the initial values, $x_1,y_1,x_2,y_2$, to obtain upper bound and cut-off value of $\vec{B}$ for any $k$ from Eq.\eqref{A1} and Eq.\eqref{A2}, respectively.

The commutation relations between momenta of both particles is obtained by Dirac quantizing the classical brackets in Eqs.\eqref{p_xjp_yk bracket}-\eqref{p_y1p_y2 bracket} as

\bea{}
[\hat{P}_{x_j},\hat{P}_{y_k}]=\frac{i}{2a^2}[(\hat{y}_k\hat{P}_{x_j}-\hat{x}_j\hat{P}_{y_k})+e(\hat{x}_j\vec{A}_{y_k}-\vec{A}_{x_j}\hat{y}_k)]
\label{p_xjp_yk com}
\eea
\bea{}
[\hat{P}_{x_1},\hat{P}_{x_2}]=\frac{i}{2a^2}[(\hat{x}_2\hat{P}_{x_1}-\hat{x}_1\hat{P}_{x_2})+e(\hat{x}\times \vec{A}_{x}).\hat{e}_3]
\label{p_x1p_y1 com}
\eea
\bea{}
[\hat{P}_{y_1},\hat{P}_{y_2}]=\frac{i}{2a^2}[(\hat{y}_2\hat{P}_{y_1}-\hat{y}_1\hat{P}_{y_2})+e(\hat{y}\times \vec{A}_{y}).\hat{e}_3]
\label{p_y1p_y2 com}
\eea
The lower bound for Eq.\eqref{p_xjp_yk com} using the same parameters as above is given by
\bea{}
(\Delta \hat{P}_{x_1})^2(\Delta \hat{P}_{y_2})^2\geq \frac{|\Lambda+ea^2\vec{A}(\hat{x}_1-\hat{y}_1)|^2}{(2k)^2a^8}\left(=3.11\times 10^{-3} \right)
\label{p_x1p_y2bound eq}
\eea
where $\Lambda=(-2+\alpha)(\hat{y}_1\hat{y}_2+\hat{x}_1\hat{x}_2)+\hat{y}_1\hat{y}_2e^{-\chi_2}+\hat{x}_1\hat{x}_2e^{-\chi_1}$, and similarly,

\bea{}
(\Delta \hat{P}_{x_2})^2(\Delta \hat{P}_{y_1})^2\geq 2.217\times 10^{-3}
\label{p_x2p_y1bound}
\eea
\bea{}
(\Delta \hat{P}_{x_1})^2(\Delta \hat{P}_{x_2})^2\geq 3.14\times 10^{-4}
\label{p_x1p_x2bound}
\eea
\bea{}
(\Delta \hat{P}_{y_1})^2(\Delta \hat{P}_{y_2})^2\geq 2.91\times 10^{-3}
\label{p_y1p_y2bound}
\eea
To comment on the lower bounds in Eqs.\eqref{p_x1p_y2bound eq}-\eqref{p_y1p_y2bound}, one needs to consider the same set of equations for the case of three entangled particles constrained on $S^1$. We assume these three particles live in three subsystems with basis $\{\psi_1,\psi_2\}$, $\{\phi_1,\phi_2\}$ and $\{\xi_1,\xi_2\}$ corresponding to the Hilbert space $\mathcal{H}_1$, $\mathcal{H}_2$ and $\mathcal{H}_3$ respectively, such that the composite state is a member of $\mathcal{H}$, where

\bea{}
\mathcal{H}=\mathcal{H}_1\otimes \mathcal{H}_2\otimes \mathcal{H}_3
\label{hilbertspace2}
\eea
 We consider a Greenberger-Horne-Zeilinger (GHZ) type state for which the $3$-tangle is maximal \cite{Lohmayer}\cite{GHZ} as
\bea{}
\psi(x_1,x_2,x_3)=\frac{1}{\sqrt{2}}(\psi_1(x_1)\phi_1(x_2)\xi_1(x_3)
+\psi_2(x_1)\phi_2(x_2)\xi_2(x_3))
\label{GHZ}
\eea
\bea{}
\psi(x_1,x_2,x_3)=\zeta e^{i(\chi_1\chi_2\chi_3)}e^{i(\chi_1+\chi_2+\chi_3)}
\label{3state}
\eea
where $\zeta=\sqrt{2}a^3$. The lower bounds corresponding to the Eqs.\eqref{p_xjp_yk com}-\eqref{p_y1p_y2 com}
are evaluated for a tripartite entangled state as
\bea{}
(\Delta \hat{P}_{x_1})^2(\Delta \hat{P}_{y_2})^2\geq \frac{|\Pi+ea^2A(\hat{x}_1-\hat{y}_1)|^2}{(2k)^2a^8} \left(= 4.9\times 10^{-4}\right)
\label{3bound1}
\eea
where $k=3$ and $\Pi=(-2+\alpha)(\hat{y}_1\hat{y}_2+\hat{x}_1\hat{x}_2)+\hat{y}_1\hat{y}_2e^{-\chi_2\chi_3}\!+\hat{x}_1\hat{x}_2e^{-\chi_1\chi_3}$.
Similarly,
\bea{}
(\Delta \hat{P}_{x_2})^2(\Delta \hat{P}_{y_1})^2\geq 9.86\times 10^{-4}
\label{3bound2}
\eea
\bea{}
(\Delta \hat{P}_{x_1})^2(\Delta \hat{P}_{x_2})^2\geq 1.4\times 10^{-4}
\label{3bound3}
\eea
\bea{}
(\Delta \hat{P}_{y_1})^2(\Delta \hat{P}_{y_2})^2\geq 1.29\times 10^{-3}
\label{3bound4}
\eea
These result are evaluated using $a^2=10$ so that $3a^2=x_1^2+x_2^2+x_3^2+y_1^2+y_2^2+y_3^2=30$ with $x_1=2$, $x_2=3.1$,$x_3=2.6$ $y_1=2.45$, $y_2=0.65$, $y_3=1.8$, charge $e=1$ and $A=0.50$.
\\
The upper bound and cutoff value of the external field for $k=3$ can be obtained from Eq.\eqref{p_xp_y com}.
The lower bound of subsystem $1$ is computed via the following equation
\bea{}
(\Delta \hat{P}_{x_1})^2(\Delta \hat{P}_{y_1})^2\geq \frac{|2+e^{-\chi_2\chi_3}+eA(x_1-y_1)|^2}{(2k)^2a^4}
\label{k3a}
\eea
and for subsystem $2$ by the following equation
\bea{}
(\Delta \hat{P}_{x_2})^2(\Delta \hat{P}_{y_2})^2\geq \frac{|2+e^{-\chi_1\chi_3}+eA(x_2-y_2)|^2}{(2k)^2a^4}
\label{k3b}
\eea
We observe that the lower bound for Eq.\eqref{k3a} and Eq.\eqref{k3b} is the same as obtained for the single particle case at $A_{cutoff}=24.5$ and $A_{upper}=2.85$ respectively. Continuing this way one can see that the computed values of $A_{upper}$ and $A_{cutoff}$ for a bipartite system with $k\geq 2$ can be inserted into a relation of the form given in Eq.\eqref{A1} and Eq.\eqref{A2}. It is easy to see that the lower bound is reduced by increasing the number of maximally entangled particles if the external field is appropriately tuned, as discussed above. Also, it is observed that for $k=3$, the upper bound on $\vec{B}$ is increased i.e.,  $\vec{B}_{\textmd{upper}}=1.80$ Tesla, and thus the system is more robust against changes in the external field. The generalization to $k>3$ for Eq.\eqref{p_xp_y com} and Eqs.\eqref{p_xjp_yk com}-\eqref{p_y1p_y2 com} is nontrivial in contrast to the Eqs.\eqref{xp com}-\eqref{y_jp_yk com}, because for $k$ particles one needs to split the global system into two blocks and the subsystems within these blocks may also be entangled with each other. One therefore needs to consider all possible bipartitions to quantify entanglement of $k$-partite system\cite{RigolinPRA}.

\section{HERMITICITY OF MOMENTUM OPERATORS }

 Even though the classical brackets are generalized to quantum commutation relations using Dirac quantization, one must note that the operator ordering problem arises in Eq.\eqref{p_xp_y com} and in Eqs.\eqref{p_xjp_yk com}-\eqref{p_y1p_y2 com}. Also, the operators appearing in these commutators must be Hermitian. So following \cite{PLA}, we try to fix the operator ordering problem by requiring the operators to be Hermitian.
The $z$-component of angular momentum is

\bea{}
\hat{L}_z^{(k)}=(\hat{x}_k\hat{P}_{y_k}-\hat{y}_k\hat{P}_{x_k})
\label{Lz}
\eea

with

\bea{}
\hat{P}_{x_k}=-\frac{\hat{y}_k}{r_k^2}\hat{L}_z^{(k)}=-\frac{1}{r_k}\sin\chi_k\hat{L}_z^{(k)}\qquad
\label{pxL}\\
\hat{P}_{y_k}=\frac{\hat{x}_k}{r_k^2}\hat{L}_z^{(k)}=\frac{1}{r_k}\cos\chi_k\hat{L}_z^{(k)}\qquad
\label{pyL}
\eea

where $k=1,2$. To find the Hermitian expression for momentum operators, we consider their expressions in the following form:

\bea{}
\hat{P}_{x_k}=-\frac{i}{r_k}\mu(\chi_k)\frac{\partial}{\partial \chi_k}+\frac{1}{r_k}\beta(\chi_k)
\label{px1}
\eea
\bea{}
\hat{P}_{y_k}=-\frac{i}{r_k}\nu(\chi_k)\frac{\partial}{\partial \chi_k}+\frac{1}{r_k}\gamma(\chi_k)
\label{py1}
\eea
These expressions are used to solve for the unknown functions $\mu(\chi_k)$, $\nu(\chi_k)$, $\beta$ and $\gamma$. Eqs.\eqref{xp com}, \eqref{yp com} and \eqref{p_xp_y com} yield information on $\mu(\chi_k)$, $\nu(\chi_k)$ and $\beta(\chi_k),\gamma(\chi_k)$ respectively.

\bea{}
\mu(\chi_k)=-\sin\chi_k
\quad,\quad
\nu(\chi_k)=\cos\chi_k
\label{mu nu}
\eea
\bea{}
\beta'=-\gamma+er_k\vec{A}\quad,\quad \gamma'=\beta-er_k\vec{A}
\label{beta' gamma'}
\eea
where prime denotes differentiation with respect to $\chi_k$. We also need the following equations of motion to be satisfied

\bea{}
i(\hat{P}_{x_k}-e\vec{A})=[\hat{x}_k,\hat{H}]\quad,\quad
i(\hat{P}_{y_k}-e\vec{A})=[\hat{y}_k,\hat{H}]
\label{eom}
\eea

The solution to these equations are given by

\bea{}
i\beta(\chi_k)\cos\chi_k+i\gamma(\chi_k)\sin\chi_k=-1+ier_k\vec{A}(\cos\chi_k+\sin\chi_k)
\label{eomresult}
\eea
where
\bea{}
\beta(\chi_k)=i\cos\chi_k+\alpha \sin\chi_k+er_k\vec{A}\label{ll}\\
\gamma(\chi_k)=i\sin\chi_k-\alpha \cos\chi_k+er_k\vec{A}
\label{beta gamma}
\eea

The momentum can thus be expressed as
\bea{}
\hat{P}_{x_k}=\frac{i}{r_k}\sin\chi_k\frac{\partial}{\partial \chi_k}+\frac{i}{r_k}\cos\chi_k+\frac{1}{r_k}\alpha \sin\chi_k+e\vec{A}
\label{px}
\eea
\bea{}
\hat{P}_{y_k}=\frac{i}{r_k}\cos\chi_k\frac{\partial}{\partial \chi_k}+\frac{i}{r_k}\sin\chi_k-\frac{1}{r_k}\alpha \cos\chi_k+e\vec{A}
\label{py}
\eea
Using the definition of an anticommutator, we can write Eqs.\eqref{px}-\eqref{py} as
\bea{}
\hat{P}_{x_k}=\frac{1}{2r_k}e^{i\alpha \chi_k}
\left\{i\frac{\partial}{\partial \chi_k},\sin\chi_k+\frac{e r_k\vec{A}}{\alpha}\right\}e^{-i\alpha\chi_k}~~~~\label{nn}\\ \hat{P}_{y_k}=\frac{1}{2r_k}e^{i\alpha\chi_k}
\left\{i\frac{\partial}{\partial \chi_k},-\cos\chi_k+\frac{er_k\vec{A}}{\alpha}\right\}e^{-i\alpha\chi_k}~~~~
\label{px anticom}
\eea
Using Weyl ordering we can write the constraint $\sigma_3=\hat{x}_i\hat{P}_{x_i}+\hat{y}_i\hat{P}_{y_i}-er_i\vec{A}$ as
\bea{}
\sigma_{3,W}=\left(-\frac{i}{2}+\hat{x}_k\hat{P}_{x_k}+\hat{y}_k\hat{P}_{y_k}\right)-er_k\vec{A}=\left(\frac{i}{2}+\hat{P}_{x_k}\hat{x}_k+\hat{P}_{y_k}\hat{y}_k\right)-er_k\vec{A}
\label{sigma3W}
\eea
Using Eq.\eqref{pxL}, we can express momentum in terms of the Weyl ordered constraint $\sigma_{3,W}$ as
\bea{}
\hat{P}_{x_k}=\frac{1}{2r_k^2}\left\{-\hat{y}_k,\hat{L}_z^{(k)}\right\}+\frac{1}{r_k^2}\hat{x}_k\sigma_{3,W}
\label{pxW}
\eea
Similarly,
\bea{}
\hat{P}_{y_k}=\frac{1}{2r_k^2}\left\{\hat{x}_k,\hat{L}_z^{(k)}\right\}+\frac{1}{r_k^2}\hat{y}_k\sigma_{3,W}
\label{pyW}
\eea
The momentum operator $\hat{P}_{x_k}$ and $\hat{P}_{y_k}$ is Hermitian when $\sigma_{3,W}$ is set to zero. One can therefore see the Weyl ordering is a prerequisite for momentum operators to be Hermitian.
Setting $\sigma_{3,W}=0$ gives $\hat{x}_k\hat{P}_{x_k}+\hat{y}_k\hat{P}_{y_k}=\frac{i}{2}+er_k\vec{A}$.
Let,
\bea{}
\hat{L}_z^{(k)}=\hat{x}_k\hat{P}_{y_k}-\hat{y}_k\hat{P}_{x_k}=-i\frac{\partial}{\partial \chi_k}-\alpha
\label{Lz op}
\eea
The Hamiltonian Eq. \eqref{H3} can be written in terms of angular momentum operator and Weyl ordered constraint $\sigma_{3,W}$ as
\bea{}
\hat{H}_T=\frac{\hat{L}_z^{(k)2}}{2r_k^2}-\frac{1}{2r_k^2}\left(\frac{i}{2}+er_k\vec{A}\right)^2-e\vec{A}(P_{r_k}-e\vec{A})+eV
\label{H4}
\eea
Solution of the Schr\"{o}dinger equation with $\!\Psi(\chi_k)=\varphi e^{i(\chi_1+\chi_2+\chi_1\chi_2)}$ leads to the following expression for energy of the system

\bea{}
E=\frac{1}{2r_k^2}[\left(\frac{1}{2}-\left[\frac{1}{2}-\bar{\alpha}\right]\right)^2\!-\bar{\alpha}(2\chi_k-1)+(\chi_k^2-\alpha)
-\frac{e\vec{A}}{2a^2}(\chi_k-\alpha')(\hat{x}_k-\hat{y}_k)+\frac{1}{2}e^2\vec{A}^2+\nonumber\\
eV]+\frac{1}{16a^2}-\frac{ie\vec{A}}{2a^2}\left(\frac{r_k}{2}+\hat{x}_k+\hat{y}_k\right)~~~
\label{energy1}
\eea
where $\bar{\alpha}=\alpha \mod 1$ and $\alpha'=\alpha-2$
\bea{}
E=E+E'
\label{energy2}
\eea
Note that the energy expression can't be split into $E_1+E_2$ because Hamiltonian is an observable of the quantum state (Eq. \eqref{polarstate}) in the Hilbert space $\mathcal{H}_1 \otimes \mathcal{H}_2$ and it is not possible to split the two particles. The  equation involves the position of both entangled particles.  If $\chi_1$ changes, $\chi_2$ also changes due to the correlation between them, such that the total momentum and total energy of the system remains conserved. Also one can see, in comparison with the quantum Hamiltonian for a bound particle (by scalar and vector potential but with no additional constraint imposed on the dynamics) given by
\bea{}
\hat{H}=\left(\frac{\hat{P}_{\chi}^2}{2r_k^2}+e^2\vec{A}^2-\frac{e\vec{A}}{r_k}\hat{P}_{\chi}+eV\right)
\label{H bound}
\eea
where $\hat{P}_{\chi}=-i\frac{\partial}{\partial \chi_k}-\alpha$.
\bea{}
E=\frac{1}{2r_k^2}\left(\frac{1}{2}-\left[\frac{1}{2}-\bar{\alpha}\right]\right)^2- \bar{\alpha}(2\chi_k-1)+(\chi_k^2-\alpha))-\frac{e\vec{A}}{r_k}\left(\chi_k-\alpha'\right)+\frac{1}{2}e^2\vec{A}^2+eV\quad\label{energy3}
\eea
the energy equation \eqref{energy1} involves an additional term $E'$. This additional term $E'$ arises due to the Dirac quantization and Hermiticity of the momentum operator. However, in comparison with \cite{PLA}, the additional energy term is not a constant and both energies are connected, since the particle positions are also involved in the additional energy term, implying that any change in particle positions simultaneously effect both the energy terms. Since $E'$ is not a constant, the term represents a shift in energy levels due to the applied magnetic field in view of similar terms arising in quantization on curved manifolds \cite{KS cheng}.
Also note that the term $\frac{e\vec{A}}{2a^2}(\chi_k-\alpha')(\hat{x}_k-\hat{y}_k)$ appearing in the energy equation is slightly different from the one obtained for the Hamiltonian in \eqref{H bound}. This is due to the requirement that $\hat{P}$ must be Hermitian and the expression for momenta involves orbit-orbit interaction between the two particles and their projection along $z$-axis is given by
\bea{}
L_z^{(1)}=\chi_2-\bar{\alpha}\quad ,\quad L_z^{(2)}=\chi_1-\bar{\alpha}
\label{Lzprojection}
\eea
The interaction between the momenta of two particles results in a total orbital angular momentum vector about $z$-axis at an angle with the direction of the external field for which the projection along $z$-axis is the scalar sum as given in the following equation.
\bea{}
L_z^{(k)}=\chi_1+\chi_2-2\bar{\alpha}
\label{Lztotal}
\eea
Hence, the total energy of the system experiences a shift due to Dirac quantization and orbit-orbit coupling between both particles.
\section{SUMMARY}
In this paper, we present a measure of uncertainty for a maximally entangled constrained bipartite system by calculating the lower bounds using Robertson relation and discussed the presence of additional energy terms for this system in connection to terms arising during quantization on curved spaces \cite{KS cheng}. In view of the work reported in \cite{Rigolin}, we emphasize that the quantum correlations also effect the lower bound when particles are constrained on $S^1$ under the influence of an external field $\vec{B}$.

We started with a classical system of two particles constrained on $S^1$ under the influence of a uniform external field $\vec{B}$. We used Dirac's method for handling degenerate Lagrangians and a set of Dirac brackets between canonical variables was obtained for this system. 
Later, we quantized this constrained classical system via Dirac's canonical quantization rule.  For this constrained quantum system, we presented a measure of uncertainty between observables by calculating the lower bounds using Robertson relation. For this, we partitioned the system into two subsystems (with $k=2$) 
and chose an inseparable (entangled) state over a composite space $\mathcal{H}_1\otimes\mathcal{H}_2$ with concurrence $C=1$, i.e., a maximally entangled state. On account of \cite{PLA}, we showed that when observables $\hat{P}$ and $\hat{x}$ act on a maximally entangled pure bipartite state, the uncertainty between the two is reduced. The reduction in uncertainty is also observed between different $\hat{P}$ and the commutation relations in Eq.\eqref{p_xjp_yk com}-\eqref{p_y1p_y2 com} are analysed for a GHZ state.

We show that the uncertainty is not only effected by the number of particles $k$, but also by the strength of the external field $\vec{B}$, and we explicitly give an expression for $\vec{A}_{\textmd{upper}}$ and $\vec{A}_{\textmd{cutoff}}$ that quantifies the measure of uncertainty as a function of the external magnetic field $\vec{B}$. We show that for $k=2$, $\vec{B}_{\textmd{upper}}=1.03$ Tesla, and for $\vec{B}\geq\vec{B}_{\textmd{upper}}$ no reduction in uncertainty is possible for atleast one subsystem and for $\vec{B}>\vec{B}_{\textmd{cutoff}}$, the bipartite entanglement is lost and there exists a direct relationship between uncertainty and the external field $\vec{B}$. It is therefore possible to appropriately tune the external field $\vec{B}$ to preserve or destroy the correlation (entanglement) between the two subsystems.
Increasing $k$ also reduces uncertainty between the observables by $1/k$ times in Eqs.\eqref{xp com}-\eqref{y_jp_yk com} as in \cite{Rigolin}, but its generalization for $k>3$ to the brackets containing different $\hat{P}$ (Eqs.\eqref{p_xjp_yk com}-\eqref{p_y1p_y2 com}) is nontrivial due to the partition of $k$-partite system into two blocks for which the number of bipartition measures increases with $k$ \cite{RigolinPRA}.


We also addressed the operator ordering problem arising due to the canonical quantization via Weyl ordering prescription and by requiring the Hermitian expressions for operators. Finally, we comment that the Schr\"{o}dinger equation for a pure bipartite entangled state contains additional energy terms arising due to the Hermiticity  of $\hat{P}$ that contains orbit-orbit interaction between the particles. This additional term $E'$ is connected with $E$ due to Eq.\eqref{polarstate}, and in view of \cite{KS cheng}, it represents a shift in the energy levels. Any changes made in particle position is reflected in the simultaneous change in both $E$ and $E'$.

\clearpage

 \bibliographystyle{unsrt}

\end{document}